\documentclass[a4paper,11pt]{article}
\usepackage{amsmath,amsthm}
\usepackage{amsfonts}
\usepackage{amssymb}
\usepackage{makeidx}
\usepackage{graphicx}
\usepackage{xcolor}
\usepackage{lmodern}
\usepackage{authblk}
\usepackage{bbm}
\usepackage{bm}
\usepackage{physics}
\usepackage{braket}
\usepackage[left=2cm,right=2cm,top=2cm,bottom=2cm]{geometry}
\usepackage{tikz}
\usepackage{verbatim}

\newtheorem{theorem}{Theorem}

\newtheorem{definition}{Definition}

\newcommand{\bs}[1]{\boldsymbol{#1}}

\newcommand{\Ran}[1]{\mathrm{Ran}{#1}}

\usepackage{color}

\title{
Toward fixed point and pulsation quantum search on graphs driven by quantum walks with in- and out-flows: a trial to the complete graph
}
\author{
Yusuke Higuchi\thanks{Department of Mathematics,
Gakushuin University, 
\small Tokyo 171-8588, Japan.}
\quad
Mohamed Sabri\thanks{Graduate School of Information Sciences, Tohoku University, Aoba, Sendai 980-8579, Japan
}
\quad
 Etsuo Segawa\thanks{Graduate School of Environment and Information Sciences, Yokohama National University, Hodogaya, Yokohama 240-8501, Japan 
 } 

 }
 \date{}

\begin{document}
\maketitle

 \noindent {\bf Abstract.} 
We treat a quantum walk model with in- and out- flows at every time step from the outside. We show that this quantum walk can find the marked vertex of the complete graph with a high probability in the stationary state. In exchange of the stability, the convergence time is estimated by $O(N\log N)$, where $N$ is the number of vertices. However until the time step $O(N)$, we show that there is a pulsation with the periodicity $O(\sqrt{N})$. We find the marked vertex with a high relative probability in this pulsation phase. This means that we have two chances to find the marked vertex with a high relative probability; the first chance visits in the pulsation phase at short time step $O(\sqrt{N})$ while the second chance visits in the stable phase after long time step $O(N\log N)$. The proofs are based on Kato's perturbation theory.  \\
 

\section{Introduction}
Quantum search driven by quantum walks on graphs is one of the interesting topics of studies on quantum walks with fruitful results (see for example \cite{Ambainis2003,Childs,Portugal} and its references). 
The time evolution operator of quantum walk is driven by a unitary operator. Since the absolute values of the eigenvalues of the unitary matrix are unit, each eigenvalue is rotated on the unit circle in the unitary time iteration. 
Then the eigenspaces of the time evolution operator which have assymptotically large overlap to the initial state with respect to the system size are dominant. 
This is the reason for the oscillation of the time course of the finding probability at the target vertex in the quantum search algorithm. Thus we need to perform the sharpshooting to the appropriate observation time, otherwise we would find the target vertex with a very small probability. 
This oscillation is known as the souffl{\'e} problem~\cite{Brassard}. 
The souffl{\'e} problem has been solved by constructing the nested time evolution which produces the convergence to a fixed point~\cite{Grover,YLC}. 

In this paper, we also consider the dynamical system which converges to a fixed point driven by a quantum walk but we set a time independent unitary evolution operator and extend the domain of the initial state from the square summable space to the uniformly bounded space. 
The initial state, which is no longer square-summable,  naturally provides the situation that the inflow penetrates the internal graph at every time step while the outflow also goes out from the internal graph. This quantum walk model is firstly introduced by \cite{FelHil1,FelHil2} as a  discrete-analogue of the stationary Schr{\"o}dinger equation~\cite{Albe}. 
In \cite{HS, MHS}, detailed aspect of the stationary state for the Szegedy walk with a constant inflow are characterized by the electric circuit. 

To see whether this quantum walk model with in- and out- flows finds the marked vertex with a high relative probability, in this paper, we try to set the complete graph with the vertex number $N$ as the internal graph and give a perturbation to the marked vertex as a first trial. We show that (i) a high relative probability at the marked vertex can be seen in the stable state; (ii) the convergence time is estimated by $O(N\log N)$; (iii) there is a pulsation with the periodicity $\Theta(\sqrt{N})$ until the time step $O(N)$. 
This means that at time step $O(\sqrt{N})$, we can find the marked vertex with a high probability. In addition, there is an insurance for missing this timing of the observation; that is, the chance to find the marked vertex with a high probability stationarily visits again after waiting for this dynamical system $O(N\log N)$. 

This paper is organized as follows. In Section~2, we set the definition of our quantum walk model. In Section~3, we give the main results. In Section~4, the proofs of the main results are shown. The proof is based on Kato's perturbation theory~\cite{Kato}.   

\section{Setting}
In this paper, we fix the internal graph as the complete graph with $N$ vertices; $K_N$. We set the semi-infinite length path and join its end vertex to each vertex of the complete graph. Such resulting infinite graph (like a hedgehog) is denoted by $\tilde{K}_N$. More precisely, 
let 
$\left\lbrace \mathbb{P}_{j}:j=1,...,N \right\rbrace$ be the set of semi-infinite length paths called the tails whose end vertex is denoted by $o(\mathbb{P}_j)$ and the vertex set of $K_N$ be $\{u_1,\dots,u_N\}$. Then the tail $\mathbb{P}_j$ is connected to the complete graph $K_N$ by identifying $o(\mathbb{P}_j)$ with $u_j\in V(K_N)$.  
Next, let us introduce the notion of the symmetric arc, which plays an important role to describe the dynamics of the quantum walk. 
The symmetric arc set induced by $E(K_N)$ and $E(\tilde{K}_N)$ by $A$ and $\tilde{A}$ respectively. For any $a\in \tilde{A}$, 
there uniquely exists the inverse arc of $a$ denoted by $\bar{a}$ in $\tilde{A}$. The origin and terminal vertices of $a\in \tilde{A}$ are denoted by $o(a)$ and $t(a)$, respectively. Then $o(a)=t(\bar{a})$ and $t(a)=o(\bar{a})$ hold. 
The degree of vertex $u\in V(\tilde{K}_N)$ is denoted by $\deg(u):=\#\{ a\in\tilde{A} \;|\; t(a)=u \}$.
Furthermore, we choose a vertex $u_*$ as the marked vertex from $V(K_N)$. 

For any countable set $\Omega$, we describe $\mathbb{C}^{\Omega}$ by the vector space whose standard basis are labeled by each element of $\Omega$. 
The total state space of our quantum walk is described by $\mathbb{C}^{\tilde{A}}$. 
The time evolution operator $U$ on $\mathbb{C}^{\tilde{A}}$ is defined as follows. 
\begin{equation}
(U\Psi)(a)=(-1)^{\boldsymbol{1}_{\{o(a)=u_*\}}(a)}\left(-\Psi(\bar{a})+\frac{2}{\deg(u)}\sum_{b\in\tilde{A}:\;t(b)=o(a)}\Psi(b)\right)  
\end{equation}
for any $a\in \tilde{A}$ and $\Psi\in \mathbb{C}^{\tilde{A}}$. 
Let $\Psi_t$ be the $t$-th iteration of the time evolution operator $U$; that is, $\Psi_{t+1}=U\psi_{t}$. 
This dynamics can be also explained as follows. 
Let $\{a_1,\dots,a_r\}\subset \tilde{A}$ be the set of all arcs whose terminal vertices are arbitrary chosen $u\in V(\tilde{K}_N)$. Note that the value $r$ takes two possibilities; if $u\in V(K_N)$, then $r=N$, while $u\notin V(K_N)$, then $r=2$. 
Then $\Psi_{t+1}$ and $\Psi_t$ locally satisfies the following recursion.   \begin{equation}\label{eq:TE2}
    \begin{bmatrix}
    \Psi_{t+1}(\bar{a}_1) \\ \vdots \\ \Psi_{t+1}(\bar{a}_r) 
    \end{bmatrix}
=\begin{cases}
-    \mathrm{Gr}(r)
    \begin{bmatrix}
    \Psi_{t}(a_1) \\ \vdots \\ \Psi_{t}(a_r) 
    \end{bmatrix} & \text{: $u=u_*$,} \\
    \\
+\mathrm{Gr}(r)
    \begin{bmatrix}
    \Psi_{t}(a_1) \\ \vdots \\ \Psi_{t}(a_r) 
    \end{bmatrix} & \text{: $u\neq u_*$.}
    \end{cases}
\end{equation}
Here $\mathrm{Gr}(r)$ is the $r$-dimensional Grover matrix; that is, $\mathrm{Gr}(r)=(2/r)J_r-I_r$, where $J_r$ and $I_r$ are the all one matrix and the identity matrix, respectively. 
This represents the dynamics that a quantum walker on the arc $b\in \tilde{A}$ is locally scattered at each vertex by the Grover matrix with the phase varied operator which is active only on the marked vertex $u_*$.  Note that if $r=2$; then the Grover matrix is reduced to the Pauli matrix $\sigma_X$. This means that the walk on the tails is the free.   
The initial state $\Psi_0\in \mathbb{C}^{\tilde{A}}$ is set so that a quantum walker penetrates $K_N$ at every time step. 
\begin{equation}
\Psi_{0}(a)= \begin{cases}
1 & \text{: $a\in \tilde{A}\setminus A$, $\mathrm{dist}(t(a),K_N)<\mathrm{dist}(o(a),K_N)$}\\ 0 & \text{: otherwise}
\end{cases} \;\;(a \in \tilde{A}),
\end{equation}
where $\mathrm{dist}(u,K_N)$ is the shortest distance between the vertex of the tail $u$ and $K_N$. 
Since the walk is free on the tails, the internal graph $K_N$ receives the inflow from the tails at every time step $t$, on the other hand, once  
a quantum walker goes out a tail, then it never go back, which can be regarded as the outflow.  
It is shown in \cite{HS} that this dynamical system converges to a fixed point; that is, there exists $\Psi_\infty\in\mathbb{C}^{\infty}$ such that
\[ \lim_{t\to\infty}\Psi_t(a)=\Psi_\infty(a) \]
for any $a\in \tilde{A}$. 
The relative probability of finding a vertex $u \in V(K_N)$ at time $t$ is given by
\[
\tilde{\nu}_{t}(u)=\sum\limits_{a \in A_{0}:t(a)=u} |\psi_{t}(a)|^{2}.
\]
Taking the normalization to $\tilde{\nu_t}$, we set
\[  \nu_t(u):=\frac{\tilde{\nu}_t(u)}{\sum_{v\in V(K_N)} \tilde{\nu}_t(v)}. \]

\section{Main results}
As a first interest, whether our quantum walker find a target vertex $u_*$ with a high probability in the long time limit. To see that, we define the following distribution 
\begin{definition}{\rm (The limit distribution)}
The normalized finding measure on $V(K_N)$ is defined by 
\[ \mu_N(u)=\lim_{t\to\infty} \nu_t(u) \]
for any $u\in V(K_N)$. 
\end{definition}
\begin{theorem}{\rm (The finding probability of the marked vertex at the fixed point)} \\
Let $\mu$ be the limit distribution of the finding probability defined as the above. Then we have 
\[ \lim_{N\to\infty}\mu_N(u_*)=1/2. \]
\end{theorem}
This theorem implies that we can find the marked vertex in the stable state with probability $1/2$ while the celebrated quantum search algorithm has the oscillation. 
Since the celebrated quantum search algorithm takes $O(\sqrt{N})$ if we measure the system at an appropriate time, 
now the next interest may be that 
{\it how does it take time for the convergence ?} 
To answer the question, let us introduce the finite time which we regard as convergence. 
\begin{definition} {\rm ($\ell^2$-mixing time of quantum walk)} 
Let $\Psi_t$  be the $t$-th iteration of the quantum walk and $\Psi_\infty$ be its stationary state.  For $\theta>0$, we set $t(\theta)$ by  
\[ t(\theta):=\min\{ s>0 \;:\; {}^\forall t>s,\;||\Psi_\infty-\Psi_t||_{K_N}<e^{-\theta} \}, \]
where for $f\in \mathbb{C}^{\tilde{A}}$, the norm $||f||_{K_N}$ is the $\ell^2$-norm with respect to the internal graph; that is,
\[ ||f||_{K_N}^2:= \sum_{a\in A:\;t(a)\in V(K_N)} |f(a)|^2. \]
\end{definition}
\begin{theorem}{\rm (The $\ell^2$-mixing time)}
Let $t(\theta)$ be the $\ell^2$-valued mixing time defined as the above. 
Then we have 
\[ t(\theta)\in \Theta(N\log N) \]
for any fixed $\theta>0$. 
\end{theorem}
Then unfortunately, in exchange for the stability, we need to wait for this quantum walk until $\Theta(N\log N)$, which means a speed down, on the contrary, this quantum walk takes more time than a classical algorithm. 
However we obtain a good news as follows. 
\begin{theorem}{\rm (Pulsation)}
Until the time step $O(N)$,  
the finding probability of the marked vertex is estimated by
\[ \nu_t(u_*)\sim \frac{1}{2}\frac{(\;1-c_t\cos[t\sqrt{2/N}] \;)^2}{1+c_t^2}, \]
where $c_t=e^{-(5/2)\cdot(t/N)}$ for large size $N$. 
In particular, at the time step $t\sim \pi \sqrt{N/2}$, the finding probability of the marked vertex is 
\[ \nu_t(u_*)\sim \frac{1}{2}\frac{(\;1+e^{-\frac{5\pi}{2\sqrt{2N}}} \;)^2}{1+e^{-\frac{5\pi}{\sqrt{2N}}}}>\frac{1}{2}. \]
\end{theorem}
This theorem implies that until the time step $O(N)$, this walk has the oscillation with the periodicity $\pi \sqrt{2N}$. 
Thus the locally maximal value of the time sequence of $\{\mu_t(u_*)\}_{t>0}$ can be observed around $O(\sqrt{N})$ and its probability is higher than $1/2$. 

\begin{figure}[htb]
  \centering
  \includegraphics[width=10cm]{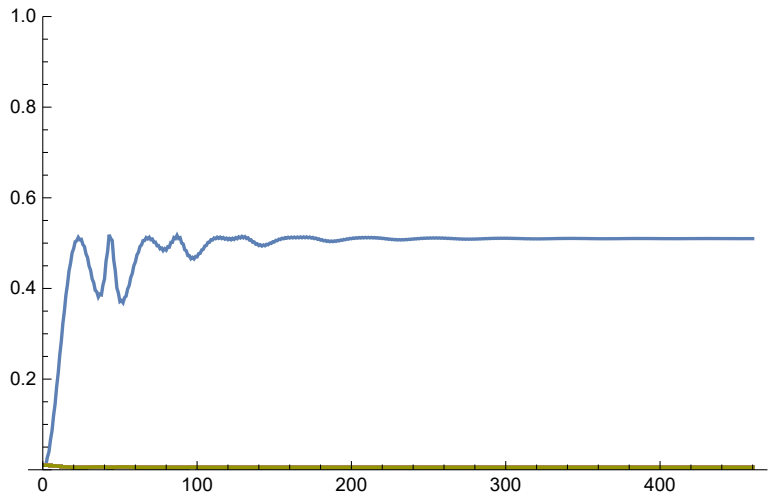}
  \caption{The time courses of the finding probability at each positions for $N=100$: The blue curve is the time course for the marked vertex. The others are those of the non-marked vertices. The convergence to $1/2$ of the blue curve derives from Theorem~1. The convergence time to $1/2$ is estimated by Theorem~2; the final time in the numerical simulation is $[N\log N]=460$. The first oscillation comes from Theorem~3; in the first half of the time course of the target vertex, the peaks can be seen with the periodicity about $\pi\sqrt{2N}\asymp 44$. }
  \label{fig:timecource}
\end{figure}

In Fig.~\ref{fig:timecource}, we give a numerical simulation of the time courses of the finding probabilities of each vertex for $N=100$. 
\section{Proofs of theorems}
\begin{proof}
\noindent{}\\
\noindent{\bf Step 1 (Reduction to an invariant subspace):} 
Let us divide the arc set $A$ into the disjoint sets $A_{+}$, $A_{-}$, $A_0$, where
\begin{align*}
    A_+ &= \{a\in A \;:\; t(a)=u_*\}, \\
    A_- &= \{a\in A \;:\; o(a)=u_*\}, \\
    A_0 &= \{a\in A \;:\; t(a),\;o(a)\neq u_*\}.
\end{align*}
From the symmetricity of the time evolution, it is easy to notice that 
the value $\psi_t(a)$ is invariant of the subsets $A_+$, $A_-$ and $A_0$. 
Thus setting $a_t,b_t,c_t$ as the representative of $\psi_t(a)\;(a\in A_+)$, $\psi_t(b)\;(b\in A_-)$ and $\psi_t(c)\;(c\in A_0)$, respectively,  we obtain 
\[
\begin{bmatrix}
a_{t+1} \\ b_{t+1} \\ c_{t+1}
\end{bmatrix}
= \begin{bmatrix}
0 & -1+2/N & 2-4/N \\ -1+2/N & 0 & 0 \\ 0 & 2/N & 1-4/N
\end{bmatrix}
\begin{bmatrix}
a_{t} \\ b_{t} \\ c_{t}
\end{bmatrix}
+\frac{2}{N}\begin{bmatrix}
1 \\ -1 \\ 1
\end{bmatrix}
\]
from the definition of our quantum walk (\ref{eq:TE2}). 
Taking the normailization, 
\[ \bs{\alpha}_t:=\begin{bmatrix} \sqrt{N-1} & 0 & 0 \\ 0 & \sqrt{N-1} & 0 \\ 0 & 0 & \sqrt{(N-1)(N-2)} \end{bmatrix}\begin{bmatrix}a_t \\ b_t \\ c_t\end{bmatrix}, \]
we obtain the following master equation:  
\begin{equation}\label{eq:TE3}
    \bs{\alpha}_{t+1}=T(\epsilon)\bs{\alpha}_t+\bs{b}_{\epsilon},\;\bs{\alpha}_0=0.
\end{equation}
Here 
\[ T(\epsilon)=\begin{bmatrix} 0 & -1+\epsilon^2 & \sqrt{2\epsilon^2(1-\epsilon^2)} \\ -1+\epsilon^2 & 0 & 0 \\ 0 & \sqrt{2\epsilon^2(1-\epsilon^2)} & 1-2\epsilon^2 \end{bmatrix} \]
and 
\[ \bs{b}_{\epsilon}=\left[\epsilon\sqrt{2-\epsilon^2},\;-\epsilon\sqrt{2-\epsilon^2},\;\sqrt{4-6\epsilon^2+2\epsilon^4} \right]^\top, \]
where $\epsilon=\sqrt{2/N}$. 
Remark that the relative probability of the marked vertex can be computed by $\nu_t(u_*)=||\bs{\alpha}_t(1)||^2$. 
Now let us consider the asymptotics for large $N$, which is equivalent to considering (\ref{eq:TE3}) for small perturbation $\epsilon$. 
\\

\noindent{\bf Step 2 (Application of Kato's perturbation theory):} The matrix $T(\epsilon)$ can be expanded as follows: 
\[ T(\epsilon)=T+ \epsilon T^{(1)}+\epsilon^2 T^{(2)}+\cdots, \]
where the non-perturbed matrix is 
\begin{align}
    T &= \begin{bmatrix} 0 & -1 & 0 \\ -1 & 0 & 0 \\ 0 & 0 & 2 \end{bmatrix},
\end{align}
which is symmetric and $T^{(1)}$ and $T^{(2)}$ are 
\[ T^{(1)}=\begin{bmatrix}
   0 & 0 & \sqrt{2} \\
    0 & 0 & 0\\
    0 & \sqrt{2} & 0
\end{bmatrix}
\text{ and } 
T^{(2)}=\begin{bmatrix}
   0 & 1 & 0 \\
   1 & 0 & 0\\
   0 & 0 & -2
\end{bmatrix}.
\]
The set of eigenvalues of $T$ is 
\begin{equation}
    \mathrm{Spec}(T)=\{-1,+1,+1\}. 
\end{equation}
The eigenvalue $1$ has the multipilicity $2$, but it is semi-simple since $T$ is a symmetric matrix. Thus by a small perturbation $\epsilon$, the eigenvalue $1$ is splitted into two parts $\lambda_{1,+}(\epsilon)$ and $\lambda_{1,-}(\epsilon)$ while the simple eigenvalue $-1$ is just perturbed which is denoted by $\lambda_{-1}(\epsilon)$. The eigenvalues of $T(\epsilon)$ can be expanded by 
\begin{align}
    \lambda_{-1}(\epsilon) &= -1+\epsilon\lambda_{-1}^{(1)}+\epsilon^2 \lambda_{-1}^{(2)}+\cdots \\
    \lambda_{1,\pm}(\epsilon) &= 1+\epsilon\lambda_{1,\pm}^{(1)}+o(\epsilon)+\cdots 
\end{align}
because $(-1)$-eigenvalue is simple while $(+1)$-eigenvalue is semi-simple. 
Now let us compute the coefficients. 
\begin{enumerate}
    \item  {\bf Expansion of $\lambda_{-1}(\epsilon)$ :} 
Since $-1$ is simple eigenvalue of the non-perturbed matrix $T$, the weighted mean of the perturbed eigenvalues of $T(\epsilon)$ around $-1$ is the perturbed eigenvalue $\lambda_{-1}(\epsilon)$ itself.  
Then following the formulas on the weighted average of the perturbed eigenvalues \cite{Kato}, let us compute the first and second terms. According to [(2.33) in Ch.~II, Sect.~2.2  \cite{Kato}], the first term and the second terms are expressed by  
\begin{align}
    \lambda_{-1}^{(1)}= \tr(T^{(1)}P_{-1}), \\
    \lambda_{-1}^{(2)}= \tr(T^{(2)}P_{-1}-T^{(1)}S_
    {-1}T^{(1)}P_{-1}),
\end{align}
where $P_{\lambda}$ is the eigenprojection of $\ker(\lambda-T)$ and $S_{\lambda}$ is the reduced resolvent at $\lambda$ ($\lambda\in\mathrm{Spec}(T)$) which can be computed by 
\[ P_{-1}=\begin{bmatrix} 1/2 & 1/2 &0 \\ 1/2 & 1/3 & 0 \\ 0 & 0 & 0 \end{bmatrix} \]
and 
\begin{align*} 
S_{-1} &= -\lim_{\xi\to -1}\sum_{\lambda \in\mathrm{Spec}(T)\setminus\{-1\}}(\xi-\lambda)^{-1}P_{\lambda}
=\frac{1}{2}(I-P_{-1})\\
& =\frac{1}{2}\begin{bmatrix}1/2 & -1/2 & 0 \\ -1/2 & 1/2 & 0 \\ 0 & 0 & 1\end{bmatrix}.
\end{align*}
Then we have 
\[ \lambda_{-1}^{(1)}=0, \text{ and } \lambda_{-1}^{(2)}=1/2. \]
As a consequence, we obtain 
\begin{equation}\label{eq:s-1}
    \lambda_{-1}(\epsilon)=-1+\frac{1}{2}\epsilon^2+O(\epsilon^3). 
\end{equation}
    \item {\bf Expansion of $\lambda_{1,\pm}(\epsilon)$:} 
Since $(+1)$-eigenvalue is semi-simple, 
but has the multiplicity, we choose a different method from the case for $\lambda_{-1}(\epsilon)$. 
Such a case, we consider 
\[\tilde{T}(\epsilon):=\frac{1}{\epsilon}(T(\epsilon)-1)P_1(\epsilon) \] 
instead of $T(\epsilon)$, which is so called the reduction process~\cite{Kato}. Here $P_1(\epsilon)$ is the total projection for  the $+1$-group.  
The new matrix $\tilde{T}(\epsilon)$ can be expanded by 
\[ \tilde{T}(\epsilon)={\tilde{T}}^{(1)}+\epsilon{\tilde{T}}^{(2)}+\cdots \]
because $+1$ is semi-simple eigenvalue of $T$. Here $\tilde{T}^{(1)}$ and $\tilde{T}^{(2)}$ are given by [(2.20) in Ch.~II, Sect. 2.2 \cite{Kato}]:
\begin{align*}
    \tilde{T}^{(1)} &= P_1T^{(1)}P_1=\frac{1}{\sqrt{2}}\begin{bmatrix}  0&0&-1\\0&0&1\\1&-1&0\end{bmatrix}, \\
    \tilde{T}^{(2)} &= P_1T^{(1)}P_1-P_1T^{(1)}P_1T^{(1)}S_1-P_1T^{(1)}S_1T^{(1)}P_1-S_1T^{(1)}P_1T^{(1)}P_1.
\end{align*} 
The coefficient $\lambda_{1,\pm}^{(1)}$ can be obtained by [(2.40) in Ch.~II, Sect.~2-3]: 
\[ \lambda_{1,\pm}^{(1)}\in \mathrm{Spec}\left( \tilde{T}^{(1)}|_{\Ran(P_1)} \right)=\{\pm i\}. \]
Since $\tilde{T}^{(1)}$ is a skew-Hermitian matrix, the eigenprojections of $\pm i$ can be directly computed by its normalized eigenvectors $\bs{v}_{1,\pm i}$ by
\[ \bs{v}_{1,\pm i}=(1/2)\;[-i,\;+i,\;\sqrt{2}]^\top. \]
Indeed we can check that 
\[ P_1=\bs{v}_{1,+i}\bs{v}_{1,+i}^*+\bs{v}_{1,-i}\bs{v}_{1,-i}^*. \]
We set $P_{1,\pm}:=v_{1,\pm}v_{1,\pm i}^*$ as a standard decomposition of $\ker(1-T)$. 
Moreover, since the eigenvalues $\pm i$ are simple, the term $o(\epsilon)$ of the expansion of $\lambda_{1,\pm}(\epsilon)$ can be expressed by  
\[o(\epsilon)=\epsilon^2\lambda_{1,\pm}^{(2)}+O(\epsilon^3) \]
and the coefficient $\lambda_{1,\pm}^{(2)}$ can be obtained by using the formula of the weighed mean of the perturbed eigenvalues again; that is, 
\[ \lambda_{1,\pm}^{(2)}=\tr(\tilde{T}^{(1)}P_{1,\pm})=-5/4. \]
After all, we conclude that
\begin{equation}\label{eq:s1}
    \lambda_{1,\pm}(\epsilon)=1\pm i\epsilon -\frac{5}{4}\epsilon^2+O(\epsilon^3). 
\end{equation}
\end{enumerate}
{\bf Step 3 (Feedback of the expansions to the expression of $\bs{\alpha}_t$): } 
By (\ref{eq:TE3}), we have 
\[ \bs{\alpha}_t=\left(I+T(\epsilon)+T^2(\epsilon)+\cdots+T^{t-1}(\epsilon)\right)\bs{b}_{\epsilon}. \]
Since $T$ is semi-simple, the perturbed matrix $T(\epsilon)$ can be decompose as 
\[ T(\epsilon)=\lambda_{-1}(\epsilon)P_1(\epsilon)+\lambda_{1,+}(\epsilon)P_{1,+}(\epsilon)+\lambda_{1,-}(\epsilon)P_{1,-}(\epsilon), \]
where $P_j(\epsilon)$ is the eigenprojection onto the each perturbed eigenvalues and can be expanded by 
\[ P_j(\epsilon)=P_j+O(\epsilon). \]
This implies 
\begin{equation}\label{eq:alpha}
    \bs{\alpha}_t=\left( \frac{1-\lambda_{-1}^t(\epsilon)}{1-\lambda_{-1}(\epsilon)}P_{-1}(\epsilon)+ \frac{1-\lambda_{1,+}^t(\epsilon)}{1-\lambda_{1,+}(\epsilon)}P_{1,+}(\epsilon)+\frac{1-\lambda_{1,-}^t(\epsilon)}{1-\lambda_{1,-}(\epsilon)}P_{1,-}(\epsilon)\right)\bs{b}_{\epsilon}. 
\end{equation} 
Inserting the expansions of $\lambda_{-1}(\epsilon)$, $\lambda_{1,\pm}(\epsilon)$ in (\ref{eq:s-1}), (\ref{eq:s1}) into (\ref{eq:alpha}), we have another expression for $\bs{\alpha}_t$ as follows: 
\begin{align}\label{eq:alpha3}
    \bs{\alpha}_t &=\frac{1}{2}\left\{1-(-1)^t e^{-\frac{\epsilon^2t}{2}\;(1+O(\epsilon))}\right\}e^{\frac{\epsilon^2}{4}(1+O(\epsilon))}\bs{b}_{\epsilon}^{(-1)} \notag \\
    &+\frac{i}{\epsilon}\left\{ 1-e^{i\epsilon t}e^{-\frac{5}{4}\epsilon^2 t(1+O(\epsilon))} \right\} e^{\frac{5}{4}i\epsilon(1+O(\epsilon))} \bs{b}_\epsilon^{(1,+)} \notag \\
    &-\frac{i}{\epsilon}\left\{ 1-e^{-i\epsilon t}e^{-\frac{5}{4}\epsilon^2 t(1+O(\epsilon))} \right\} e^{-\frac{5}{4}i\epsilon(1+O(\epsilon))} \bs{b}_\epsilon^{(1,-)}.
\end{align}
Here $\bs{b}_\epsilon^{(j)}:=P_{j}(\epsilon)\bs{b}_\epsilon$.
We notice that $P_{-1}\bs{b}_{0}=\bs{0}$, which implies that there exists $\bs{b}'$ such that $\bs{b}_{\epsilon}^{(-1)}=\epsilon(\bs{b}'+O(\epsilon))$. 
\\

\noindent{\bf Step~4 (Finishing the proofs):} \\
\noindent \underline{Proof of Theorem~1.}
Taking $t\to\infty$ in (\ref{eq:alpha}) and inserting the expansions of $\lambda_{-1}(\epsilon)$ and $\lambda_{1,\pm}(\epsilon)$ in (\ref{eq:s-1}) and (\ref{eq:s1}), we have 
\begin{align*}
    \bs{\alpha}_{\infty} &:=\lim_{t\to\infty}\bs{\alpha}_t \\
    &= \frac{1}{2}e^{\epsilon^2(1+O(\epsilon))/4}P_{-1}(\epsilon)\bs{b}_{\epsilon}+\frac{i}{\epsilon}e^{5i\epsilon(1+O(\epsilon))/4}P_{1,+}(\epsilon)\bs{b}_{\epsilon}-\frac{i}{\epsilon}e^{-5i\epsilon(1+O(\epsilon))/4}P_{1,-}(\epsilon)\bs{b}_{\epsilon}.
\end{align*}
Then 
\begin{align*}
\lim_{\epsilon\to 0}\epsilon \bs{\alpha}_\infty &= 
\lim_{\epsilon\to 0} (ie^{5i\epsilon/4}P_{1,+}(\epsilon)\bs{b}_{\epsilon}-ie^{-5i\epsilon/4}P_{1,-}(\epsilon)\bs{b}_{\epsilon}) \\
&= \lim_{\epsilon\to 0} 2\mathrm{Re}\left( ie^{5i\epsilon/4}P_{1,+}\bs{b}_{\epsilon} \right) \\
&= -2\mathrm{Im}(P_{1,+}\bs{b}_0) \\
&= \frac{-1}{\sqrt{2}}\begin{bmatrix}-i \\ i \\0 \end{bmatrix}. 
\end{align*}
Thus the finding probability at the target vertex $u_*$ is $1/2$ because the relative probability is given by  $|\bs{\alpha}_\infty(1)|^2$. \\

\noindent \underline{Proof of Theorem~2.}
By (\ref{eq:alpha3}), we have 
\begin{align}\label{eq:dif}
    ||\bs{\alpha}_\infty-\bs{\alpha}_t|| &= \frac{e^{-\frac{\epsilon^2t}{2}(1+o(1))}}{\epsilon}\; \bigg|\bigg|\frac{\epsilon^2}{2}(-1)^t(\bs{b}'+o(1))+ie^{-\frac{3}{4}\epsilon^2t}(\bs{b}_0^{(1,+)}+o(1))-ie^{-\frac{3}{4}\epsilon^2t}(\bs{b}_0^{(1,-)}+o(1))\bigg|\bigg|.
\end{align}
There is a conflict between $\epsilon^2$ and $e^{-3\epsilon^2t/4}$ in the RHS. 
Note that $\epsilon^2\ll e^{-3\epsilon^2t/4}$ if and only if $t< 8|\log \epsilon|/(3\epsilon^2)$. 
Let us see the lower bound of $t$ such that $||\bs{\alpha}_\infty-\bs{\alpha}_t||<e^{-\theta}$ if $t<8|\log \epsilon|/(3\epsilon^2)$. 
By (\ref{eq:dif}), there exists $m>0$ such that 
    \begin{align}
        ||\bs{\alpha}_\infty-\bs{\alpha}_t|| &= (m+o(1)) \frac{e^{-\frac{5\epsilon^2t}{4}(1+o(1))}}{\epsilon}. 
    \end{align}
    Then solving 
    \[  (m+o(1)) \frac{e^{-\frac{5\epsilon^2t}{4}(1+o(1))}}{\epsilon}<e^{-\theta}, \]
    we obtain the lower bound of $t$ by  
    \[t>(\theta+\log(m+o(1))+|\log \epsilon|)\frac{4}{5\epsilon^2(1+o(1))}. \]
Thus the lower bound of such a $t$ must belong to at least $\Theta(|\log \epsilon|/\epsilon^2)$. 
On the other hand, if $t\geq 8|\log \epsilon|/(3\epsilon)^2$, then (\ref{eq:dif}) is rewritten by 
\begin{align}
    ||\bs{\alpha}_\infty-\bs{\alpha}_t||&= (m'+o(1)) \epsilon e^{-\frac{\epsilon^2 t}{2}(1+o(1))},
\end{align}
which implies 
\[ t>(\theta+\log(m'+o(1))-|\log \epsilon|)\frac{2}{\epsilon^2 (1+o(1))}. \]
Since $t>8|\log \epsilon|/(3\epsilon)^2$, this inequality is satisfied for any fixed $\theta$ if $\epsilon$ is sufficiently small. 
Therefore we can conclude that $t(\theta)\in \Theta(|\log \epsilon|/\epsilon^2)=\Theta(N\log N)$. 
%
\\

\noindent \underline{Proof of Theorem~3.}
Since $\bs{b}_\epsilon^{(-1)}\in O(\epsilon)$, if $t\ll 1/\epsilon^2$, the second and third terms in (\ref{eq:alpha}) are main terms; that is, 
\[ \bs{\alpha}_t\sim \frac{i}{\epsilon}(1-e^{i\epsilon t} e^{-\frac{5}{4}\epsilon t^2})\bs{b}_\epsilon^{(1,+)}-\frac{i}{\epsilon}(1-e^{-i\epsilon t} e^{-\frac{5}{4}\epsilon t^2})\bs{b}_\epsilon^{(1,-)} \]
for sufficiently small $\epsilon$. 
Note that 
\begin{align*}
\bs{b}_{\epsilon}^{(1,+)} &= P_{1,+}\bs{b}_{\epsilon} 
= \langle \bs{v}_{1,i},\bs{b}_0  \rangle\bs{v}_{1,i}+O(\epsilon) \\
&=[-i/\sqrt{2},i/\sqrt{2},1]^\top+O(\epsilon).\\
\bs{b}_{\epsilon}^{(1,-)} &= P_{1,-}\bs{b}_{\epsilon} 
= \langle \bs{v}_{1,-i},\bs{b}_0  \rangle\bs{v}_{1,-i}+O(\epsilon) \\
&=[i/\sqrt{2},-i/\sqrt{2},1]^\top+O(\epsilon).
\end{align*}
Therefore until $t\in O(1/\epsilon)$, 
we have 
\begin{align*}
    \bs{\alpha}_t\sim \frac{i}{\epsilon} \left[\;-\sqrt{2}(1-c_t\cos\epsilon t),\; \sqrt{2}(1+c_t\cos\epsilon t),\;2\sin\epsilon t\;\right]^\top,
\end{align*}
where $c_t=e^{-(5/4)\epsilon^2 t}$. 
Then the normalized constant of the relative probability can be computed by $4(1+c_t^2)$. Then we have 
\[ \nu_t(u_*)\sim \frac{2(1-c_t\cos\epsilon t)^2}{4(1+c_t^2)},  \]
for $t\in O(1/\epsilon)$. 
In particular, if $t=\pi/\epsilon$, then the finding probability at the marked vertex takes the local maximal value 
\[ \nu_t(u_*)=\frac{1}{2} \frac{(1+e^{-5\pi/(2\sqrt{2N})})^2}{1+e^{-5\pi/\sqrt{2N}}}>1/2. \]
\end{proof}
\section{Summary}
In this paper, we considered a quantum search driven by quantum walk with in- and out-flows. 
We mark one-vertex in the complete graph. 
The internal graph receives the inflow and releases the out flow at every time step. 
This quantum walk model converges to a fixed point.
We find that there are two phases of the time evolution of this quantum walk: the first phase is the pulsation phase until $O(N)$ and the send phase is the stationary phase after $O(N\log N)$. 
We show that (i) the high probability of the marked vertex in the stationary phase (Theorem~1); (ii) the mixing time is estimated by $O(N\log N)$ (Theorem~2); (iii) until $O(N)$, there is a pulsation with the periodicity $O(\sqrt{N})$ and we find the marked vertex with a high probability in this pulsation phase (Theorem~3). 
By extending this model to general graphs, to classify the graphs where such an aspect of a kind  of a phase transition appear is one of the interesting future's problems. 
\vskip\baselineskip
\noindent{\bf Acknowledgments}: M.S. acknowledges financial supports from JST SPRING (Grant No.~JPMJSP2114). Yu.H. acknowledges financial supports from the Grant-in-Aid of
Scientific Research (C) Japan Society for the Promotion of Science (Grant No.~18K03401). 
E.S. acknowledges financial supports from the Grant-in-Aid of
Scientific Research (C) Japan Society for the Promotion of Science (Grant No.~19K03616) and Research Origin for Dressed Photon.
The authors would like to thank prof. Hajime Tanaka for the invaluable discussions and the comments which were very useful to carry out this study.

\end{document}